%% file: paper.tex
\documentclass{sig-alternate-2013}
\usepackage[usenames,dvipsnames]{color}
\usepackage{epstopdf}
\usepackage{longtable}
\usepackage{tabularx}
\usepackage{afterpage}
\usepackage{booktabs}

\include{definitions}

\usepackage{xcolor}
\usepackage{enumitem}
\usepackage{amsfonts}
\usepackage{scalerel}
\usepackage{tikz}
\usepackage{flushend}

\begin{document}

\title{P-score: A Publication-based Metric\\ for Academic Productivity}

\numberofauthors{5} 
\author{
\alignauthor
Sabir Ribas\\
       \affaddr{CS Dept, UFMG}\\
       \affaddr{Belo Horizonte, Brazil}\\
       \email{sabir@dcc.ufmg.br} \\
\alignauthor
Berthier Ribeiro-Neto\\
       \affaddr{CS Dept, UFMG \& Google Inc}\\
       \affaddr{Belo Horizonte, Brazil}\\
       \email{berthier@dcc.ufmg.br} \\
\alignauthor
Edmundo de Souza e Silva\\
       \affaddr{COPPE, UFRJ}\\
       \affaddr{Rio de Janeiro, Brazil}\\
       \email{edmundo@land.ufrj.br} \\
\and  
\alignauthor Alberto Ueda\\
       \affaddr{CS Dept, UFMG}\\
       \affaddr{Belo Horizonte, Brazil}\\
       \email{ueda@dcc.ufmg.br}
\alignauthor Nivio Ziviani\\
       \affaddr{CS Dept, UFMG \& Zunnit Tech}\\
       \affaddr{Belo Horizonte, Brazil}\\
       \email{nivio@dcc.ufmg.br}
}

\date{}

\maketitle

\begin{abstract}
In this work we propose a metric to assess academic productivity based on publication outputs. We are interested in knowing how well a research group in an area of knowledge is doing relatively to a pre-selected set of reference groups, where each group is composed by academics or researchers. 
To assess academic productivity we propose a new metric, which we call P-score.
%
%
%
Our metric P-score assigns weights to venues using only the publication patterns of selected reference groups. 
%
%
This implies that P-score does not depend on citation-data and thus, that it is simpler to compute particularly in contexts in which citation data is not easily available. Also, 
preliminary experiments suggest that P-score preserves strong correlation with citation-based metrics.
\end{abstract}




\section{Introduction}
\label{sec:introduction}

The assessment of academic productivity usually involves the association of metrics with the researchers or groups of researchers one wants to evaluate. Funding agencies, university officials, and department chairs are examples of entities interested in these metrics, as these have application in a variety of practical situations. 
%
%
There are also cases in which one needs to compare researchers working on a same sub-area of knowledge, some examples are finding review peers, constructing program committees or compiling teams for grants.

Today, the most reliable and complete way to compare researchers is by compiling information on their academic output such as number of publications, citation based metrics, number of undergraduate and graduate students under supervision, number of advised masters and PhD theses, and participation in conferences and in technical committees. Some councils also use extensive surveys to compile qualitative information on features associated with the programs. 

However, as compiling this information is not a simple task and takes a long time, it is a common procedure to use just citation data to gain quick insights into the productivity of research groups and academics.
But, given that compiling citation counts requires access to the contents of a large pool of publications, which is not always available, new and complementary metrics, such as P-score, are a necessity. 
%

The notion of academic productivity is intrinsically associated with the notion of reputation.
And although the concept of reputation lacks on definition, we can see it as a simple property of an individual or group which measures their academic impact in the world and which we can associate metrics with. 
To measure the reputation of researchers, it is a common procedure to use the publication venues they publish in. Higher the impact of a venue, higher is considered the reputation of the researchers who publish in it. We use this idea of transferring reputation through publications to introduce a new metric called P-score. 

\section{The P-score Approach}
\label{sec:method}

The question we address in this work is:
\textit{How to model research groups, researchers and venues to capture the notion of relevance or importance of each, using only information about (i) the relationship of groups and members and (ii) the list of publication records of each member, without using paper contents or citation counts?}
Working with this question, we emerged with a metric, which we call P-Score. 

\subsection{Overview and Assumptions}


The basic idea of P-score 
is to associate a reputation with publication venues based on the publication patterns of 
a set of {\em reference groups} of researchers in a given area or sub-area of knowledge. 
For now we consider
that it is possible to select such references, even if it might be controversial. 

%
We assume that the reputation of a research group is strongly influenced by the reputation of its members, 
which is largely dependent on their publication records. 
P-score is based on the following assumptions:
\begin{enumerate} 
\item A researcher or a group member conveys reputation to a venue proportionally to its own reputation.
\item The reputation of a researcher is proportional
to the reputation of the venues in which he/she publishes.
\end{enumerate}
Once a reference group in a given area is selected, the reputation of members in this group
is transferred to the venues.
A Markov chain model can then be built from these ideas.

\subsection{Notation and Publication Counts}

Before developing the model, we introduce some notation.
Table \ref{t:notation} summarizes the notation and definitions used in this work. 
We use $\omega$ 
and $j$ as indexes for research groups 
and the venues where they publish, respectively.
The research groups used as reputation sources are referred to jointly as the {\em reference groups}. 
Consider a chosen set $\ST$ of reference groups, and let $T$ be its cardinality.
Let $\SV$ be the set of all venues $\mv_j$ where the groups in $\ST$ publish, and $V$ the total number of
venues in the set $\SV$. 
Members of research group
$\omega$ publish in subset $\SV_\omega \subseteq \SV$ with cardinality 
$V_\omega = | \SV_\omega |$.

\begin{table}[h] \small
\caption{Notation}
\label{t:notation}
\begin{tabularx}{0.47\textwidth}{|c|X|}
\hline
$\ST$	&set of reference groups \\
$T$	        &cardinality of $\ST$ \\
$\omega$	&a research group in $\ST$ \\
$\SV$	&set of venues where the researchers in $\ST$ publish  \\
$V$	&cardinality of $\SV$ \\
$\SV_\omega$	&set of venues where the researchers of group $\omega$ publish  \\
$V_{\omega}$	&cardinality of $\SV_\omega$ \\
$\mbox{v}_j$	&the $j^{th}$ venue where members of a group in $\ST$ publishes at \\
$N(\omega, \mbox{v}_j)$	&total number of distinct papers published by group $\omega$ in venue $\mv_j$ \\
$N(\mbox{v}_j)$	&total number of papers published in venue $\mv_j$ \\
$N(w)$	&total number of publications of group $\omega$ \\
$D(v_j)$	& number of distinct authors publishing in venue $v_j$ \\ 
$\gamma_{\omega}$	&reputation of group $\omega \in \ST$ \\
$\nu_j$	&reputation of venue $\mv_j \in \SV$ \\ 
%
\hline
\end{tabularx}
\end{table}

We define a function $N$ that counts the papers published by research groups and the papers published at venues. 
Let $N(\omega, \mbox{v}_j)$ be the total number of {\em distinct papers} published by research group $\omega$ in venue $\mv_j$ 
and let $N(\mbox{v}_j)$ and $N(w)$ be the total number of papers published in 
venue $\mv_j$ and
the total number of publications of group $\omega$ during the observation period, respectively.
That is:
\begin{eqnarray*}
N(w) & = & \sum_{j=1}^{V} N(\omega, \mbox{v}_j) \\
N(\mbox{v}_j) & = & \sum_{w=1}^{T} N(\omega, \mbox{v}_j) 
\end{eqnarray*}

\subsection{A Markov Model of Reputation}

From Assumption 1, the reputation of reference group $w$ is defined as:
\begin{eqnarray}
\gamma_w & = & \sum_{j=1}^V \nu_j \times \alpha_{wj} \label{eq:rep-dep}
\end{eqnarray}
where 
\begin{equation}
\label{eq:alpha}
\alpha_{wj} = \frac {N(\omega, \mbox{v}_j)} {N(\mbox{v}_j)}
\end{equation}
is the fraction of publications of venue $\mv_j$ that are from research group $\omega$ 
and $V$ is the number of venues. 

Let $D(v_j)$ be the number of distinct authors that publish in venue $v_j$
and $T$ the number of reference groups. 
From Assumption 2, the reputation of venue $v_j$ is defined as:
\begin{equation}
\label{eq:rep-ven}
\nu_j = \sum_{w=1}^T \gamma_w \times \beta_{wj} 
\end{equation}
where
\begin{equation}
\label{eq:beta}
\beta_{wj} = d \times \frac {N(\omega, \mbox{v}_j)} {N(w)} + (1-d) \times \frac {D(\mbox{v}_j)} {\sum_k {D(\mbox{v}_k)}}
\end{equation}
combines the fraction of publications of group $\omega$ that are from venue $v_j$ and the fraction of distinct authors that publish in $v_j$. The intuition for this formulation is venues that receive publications from a small set of authors are most likely to have lower reputation, e.g. local workshops may receive a large amount of publications but the total number of distinct authors tend to be small. The parameter $d$ $(0\le d\le 1)$ controls the relative importance between the volume of publications that $v_j$ receives from a group $\omega$ and the total number of authors publishing there.

If $d=1$ then the reputation of the publication venues is totally derived from the reference groups. If $d=0$ then the reputation of the publication venues is totally derived from the amount of distinct authors (from reference groups or not) publishing there. We noticed that varying $d$ does have an impact on venue weights. 

Let $\bP$ be a $(T+V) \times (T+V)$ square matrix such that element $p_{mn} = 0$ if 
either $m, n \leq T$ or $m, n \geq T$.
In addition, $p_{mn} = \beta_{m,n-T}$ for $m \leq T, n > T$ and   
$p_{mn} = \alpha_{m-T,n}$ for $m > T, n \leq T$.
Note that, since $\sum_{w=1}^{T} \alpha_{wj} = 1$ for all $ 1 \leq j \le V$ and 
$\sum_{j=1}^{V} \beta_{wj} = 1$ for all $1 \leq w \leq T$ then
$\bP$ defines a Markov chain.
In addition, the Markov chain is periodic and has the following structure:
\[
\bP 
=
\left[
\begin{array}{c | c}
\bzr      &\bP_{12} \\
\hline
\bP_{21}  &\bzr    \\
\end{array}
\right]
=
\left[
\begin{array}{ c@{\hspace{0.29cm}}c@{\hspace{0.29cm}}c | c@{\hspace{0.29cm}}c@{\hspace{0.29cm}}c}
0            &\ldots    &0           &\beta_{11}   &\ldots   &\beta_{1V} \\
\vdots       &\ddots    &\vdots      &\vdots       &\ddots   &\vdots    \\
0            &\ldots    &0           &\beta_{T1}   &\ldots   &\beta_{TV} \\
\hline 
\alpha_{11}  &\ldots    &\alpha_{T1} &0            &\ldots   &0  \\
\vdots       &\ddots    &\vdots      &\vdots       &\ddots   &\vdots \\
\alpha_{1V}  &\ldots    &\alpha_{TV} &0            &\ldots   &0 
\end{array}
\right]
\]
\noindent 
From decomposition theory, see \cite{meyer89}, we can obtain values for 
ranking the reference groups by solving:
\begin{equation}
\label{eq:top-rank}
\bgamma = \bgamma \bP^\prime
\end{equation}
where $\bP^\prime = \bP_{12} \times \bP_{21}$ is a stochastic matrix
and $\bgamma = \langle \gamma_1, \ldots, \gamma_T \rangle$.
Note that matrix $\bP^\prime$ has dimension $T \times T$ only and can be easily solved by
standard Markov chain techniques such as the GTH algorithm \cite{GTH85}.
Then, from Equation (\ref{eq:rep-dep}) we obtain the reputation of all venues where the reference groups publish.
\begin{equation}
\label{eq:venue-rank}
\bnu = \bgamma \times \bP_{12}
\end{equation}

This vector of venue P-scores can be used to rank authors (or even research groups) one want to compare. But, before continue the development of the P-score model, it is convenient to discuss a small example to illustrate the notation.

\subsection{Example}

\begin{figure}[h]
   \centerline{\includegraphics[width=8cm]{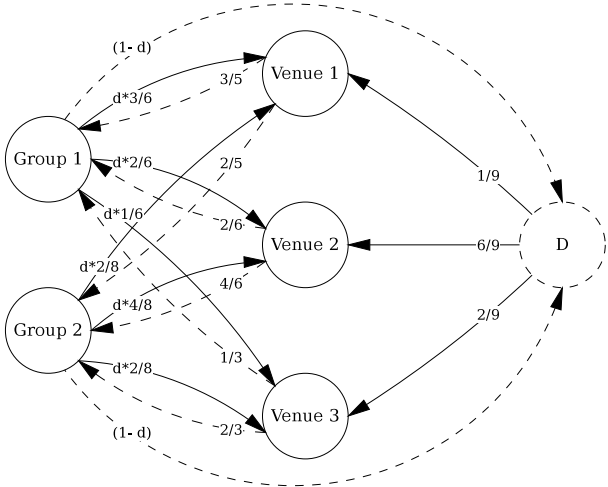}}
   \caption{Markov chain for a small example with 2 research groups and 3 venues.}
   \label{fig:ex1-MC}
\end{figure}

Figure~\ref{fig:ex1-MC} illustrates the Markov chain associated with a 
small example composed of two reference research groups and three publication venues. 
In this example, 
faculty members of Group 1 published
a total of six papers, three of which 
in venue $v_1$, two in venue $v_2$, and one in venue $v_3$.
Venue $v_1$ got also two papers from faculty of Group 2.
Since venue $v_1$ has a total of five papers from Groups 1 and 2, 
its reputation is distributed to the two groups proportionally to the
number of papers from each.
The remaining publication patterns are shown in the figure. 

Consider also that we have the number of authors that publish in each venue as an additional information.
In our example, assume that venues 1, 2 and 3 receive publications from 10, 60 and 20 distinct authors, respectively.
Our intuition is that venues with a larger number of distinct authors are better than venues with a
small number of authors (i.e., we penalize venues that are recognized by a few authors). We refer to this effect as the {\em publication breadth} of the venue.
This information is modeled through the \textit{dangling node} $D$ and the parameter $d \in [0,1]$, 
which we use to balance the relative importance of publication volume and publication breadth 
in the model.
If $d=1$ then only publication volume is considered. If $d=0$ then only publication breadth is considered.
For effect of illustration, consider that $d=1/3$ in our small example of Figure~\ref{fig:ex1-MC}.
Then, we can write an stochastic transition matrix $P$ as follows:
\[
P =
\left[
\begin{array}{ @{\hspace{0.45cm}}c@{\hspace{0.45cm}}c@{\hspace{0.45cm}} | @{\hspace{0.45cm}}c@{\hspace{0.45cm}}c@{\hspace{0.45cm}}c@{\hspace{0.45cm}} }
  &  &  &  & \\
0  &0  &\frac{1}{3}\frac{3}{6}+\frac{2}{3}\frac{1}{9}  &\frac{1}{3}\frac{2}{6}+\frac{2}{3}\frac{6}{9}  &\frac{1}{3}\frac{1}{6}+\frac{2}{3}\frac{2}{9} \\
  &  &  &  & \\
0  &0  &\frac{1}{3}\frac{2}{8}+\frac{2}{3}\frac{1}{9}  &\frac{1}{3}\frac{4}{8}+\frac{2}{3}\frac{6}{9}  &\frac{1}{3}\frac{2}{8}+\frac{2}{3}\frac{2}{9} \\
  &  &  &  & \\
\hline
  &  &  &  & \\
\frac{3}{5}          &\frac{2}{5}           &0         &0           &0       \\
  &  &  &  & \\
\frac{2}{6}          &\frac{4}{6}           &0         &0           &0       \\
  &  &  &  & \\
\frac{1}{3}          &\frac{2}{3}           &0         &0           &0       \\
  &  &  &  & \\
\end{array}
\right]
\]

Given $P$, we can compute the steady state probabilities associated with each venue to obtain 
the vector $\nu$ of all venues:
\begin{eqnarray}
\nu & = \langle 0.189, 0.590, 0.221 \rangle \\
    & = \langle 0.320, 1.000, 0.375 \rangle
\end{eqnarray}
%
%
%
The values in vector $\nu$ are the venue P-scores. In our example, venue $\mbox{v}_2$ has the highest P-score, followed by $\mbox{v}_3$, and then by $\mbox{v}_1$. We remark that the individual values give the relative importance of each venue with respect to $v_2$.

\subsection{Comparing Authors}

Once the vector $\nu$ of venue P-scores has been computed, we can easily compute a rank $\cal{R}$ for each author $a$ in a set of authors $A$ we want to compare as: 
\begin{equation}
\label{eq:p-score}
\mbox{$\cal{R}$}(a \in A) = \frac{S_a}{\max_{i \in A} \{ S_{i} \} } 
\end{equation}
where 
$S_a$ $(a \in A)$ is a weighted sum of P-scores associated with author $a$ in set $A$,
computed as: 
\begin{equation}
\label{eq:gamma_chi}
S_a = \sum_{j=1}^V \nu_j \times N(a,v_j)
\end{equation}
where 
$\nu_j$ is the weight (or P-score value) of venue $v_j$ according to $\nu$ and 
$N(a,v_j)$ is the total number of publications from author $a$ in venue $v_j$.

\section{Discussion}
\label{sec:discussion}

We have proposed an metric to assess academic productivity, which we call P-score, given it is based just on the {\em publication} patterns of research groups.
%
%
The basic idea of P-score is to associate a reputation with publication venues based on the publication patterns of reference groups, composed by researchers, in a given area of knowledge. 
%
%
Although the choice of reference groups can be made by using available citation data, the P-score metric itself does not depend on citation data. It uses just publication records of researchers and research groups, i.e. the papers and the venues where they published in. 
%
%
Preliminary experiments suggest that results have strong correlation with citation-based metrics and yet, have some complementarity to them, something we are further investigating.

\section*{ACKNOWLEDGEMENTS}
This work was partially sponsored by the Brazilian National Institute of Science and Technology for the Web (MCT/CNPq 573871/2008-6) and the authors' individual grants and scholarships from CNPq, FAPEMIG and FAPERJ.

\bibliographystyle{abbrv}
\bibliography{paper,bib-edm-short}

\end{document}

%% file: definitions.tex
\newcommand{\mv}{ {\mbox{v}} }

\newcommand{\ST}{ {\cal T} }

\newcommand{\SV}{ {\cal V} }

\newcommand{\bP}{ {\bf P} }

\newcommand{\bgamma}{ {\mbox{\boldmath $\gamma$}} }

\newcommand{\bnu}{ {\mbox{\boldmath $\nu$}} }

\newcommand{\bzr}{ {\bf 0} }